\documentclass[aps,prd,onecolumn,12pt,preprintnumbers]{revtex4-1}

\usepackage{hyperref}

\usepackage{float}

\usepackage{tikz}

\usepackage{graphicx}
\usepackage{epstopdf}

\begin{document}

\preprint{HDP: 21 -- 01}

\title{Pickers' Guide to {\it Acoustics of the Banjo, parts I and II}}

\author{David Politzer$^*$, Jim Woodhouse$^\dagger$, and Hossein Mansour$^+$}

\date{April 2, 2021}

\begin{figure}[h!]
\includegraphics[width=4.8in]{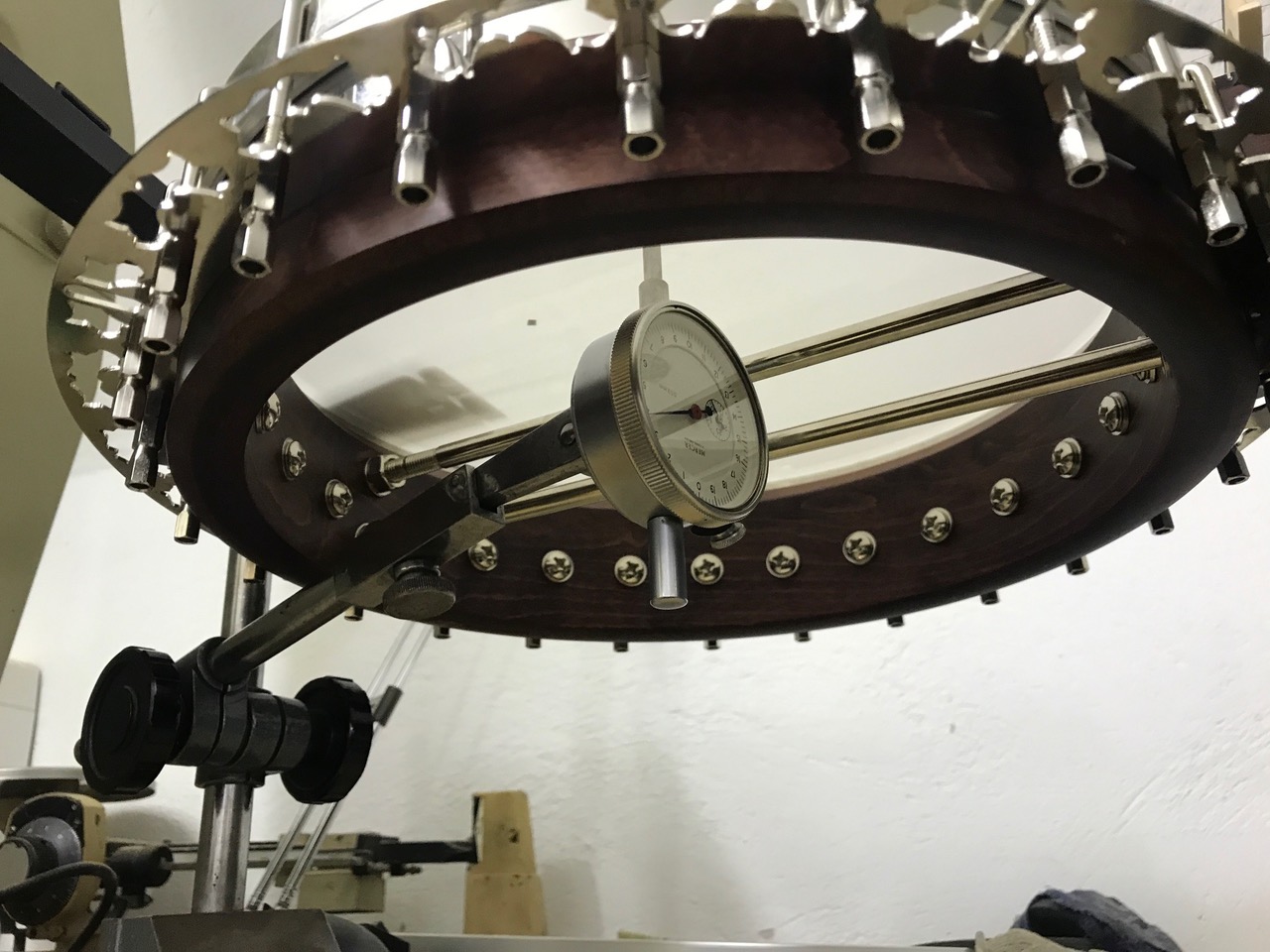}
\end{figure}

\begin{abstract}

The results of an investigation of fundamental banjo acoustics and physics are now readily available.\cite{acoustics, euphonics}.  The papers describe measurements and calculations which demonstrate the extent to which the sound of a banjo can be represented by linear modeling of the basic parts common to the banjo family: strings, drum head, light floating bridge, and break angle. This note summarizes the conclusions of that work and outlines some of the physics involved.  While the scrutinized banjo behaviors are well-known to discerning builders and pickers, the focus here is how these arise from the physics.  A practical application (discussed in an addendum) concerns head tap-tuning: what's going on, how it works, and why some people and some tuners can't get it right.

\bigskip

\bigskip\noindent $^*$452-48 Caltech, Pasadena CA 91125; $^\dagger$Cambridge University Engineering Department, Trumpington Street, Cambridge CB2 1PZ; $^+$Dassault Syst\`emes -- SIMULIA, 5005 Wateridge Vista Dr., San Diego, CA 92121
\end{abstract}

\maketitle{ {\centerline{\large \bf  Pickers' Guide to {\it Acoustics of the Banjo, parts I and II}}}

\bigskip  

\section{Introduction}

What happens when you use a drum head for the soundboard of a plucked instrument?  You get a banjo --- or one of its many cousins from cultures around the world.  Their sounds share features that distinguish them from instruments that are similarly strung and played but use thin wood (or some such material) in place of the drum head.
  
``What happens...?" can also mean: What's going on from a physics perspective?  How does it work?  

In all acoustic string instruments, string vibrations activate the soundboard. The soundboard vibrations disturb the surrounding air, and that turns into the sound we hear.  The simplest distinction of the banjo from, say, an identically strung guitar of comparable size, is that the head and bridge are much lighter than the analogous parts on the guitar.  Energy is transferred faster by the strings to the banjo head, and it moves more.  So, the banjo sound is louder at first and decays faster.  But there are many more consequences of the drum-as-soundboard.  (To hear an example, you can go to 
\noindent \href{http://www.its.caltech.edu/~politzer/pickers-guide/four-pluck-comparison.mp3}{http://www.its.caltech.edu/{\url{~}}politzer/pickers-guide/four-pluck-comparison.mp3} 
or simply \href{http://www.its.caltech.edu/~politzer/pickers-guide/four-pluck-comparison.mp3}{click here}
\noindent for gentle plucks on a typical banjo and a guitar, at the same place along equal scale-length open steel strings.)  
Using acoustic guitar again for comparison, well-studied guitar physics and acoustics imply that, if it were physically possible to build a wood soundboard and bridge as light as a banjo's, it would still not sound like a banjo.

Physics and structural engineering deal with numbers and equations.  Ref.~\cite{acoustics} translates banjo into a form appropriate for that kind of analysis.  The approach and tools are mostly ones long used to study stringed instruments.  There is also a body of previous research that has specifically addressed drums.  However, it hadn't all been put together to study drum-topped string instruments.  In addition, yet other physics from other contexts enters as well.  Guitar and violin, which are both more widely familiar and have been intensively studied, provide useful comparisons.  Ref.~\cite{acoustics} has links to sound files at an on-line science of musical instruments resource, {\it Euphonics}\cite{euphonics}, put together by author J.W.  These are sounds synthesized using the physics models developed in ref.~\cite{acoustics}.  You can judge for yourself how well straightforward linear modeling of strings, bridge, and head account for banjo sound.  Furthermore, the {\it Euphonics} sound links include variations of the relevant parameters, such as bridge mass and break angle.  The agreement of the variation in modeled sounds with actual players' experience is yet further confirmation that the basics have, indeed, been captured by the modeling.  Sketched  below are some of the physics details of that work.

Strings, floating bridge, and drum-like head are the elements common to all instruments with banjo sound.  To be sure, there are many other features whose design variations contribute discernible sound differences.\cite{coupled}  To understand those, knowledge of the basics can't but help.  There are also certainly non-linear effects, even in the basic elements, that can be heard if one listens closely.  

\section{energy \& dissipation}

On an acoustic guitar (or any wood topped instrument), sound carries away only a small fraction of the energy that is put in by a pluck.  Most of the energy is dissipated into heat.  The dominance of dissipation into heat is frustrating if you want a simple perspective on what's going on.  Those losses are functions of frequency that cannot be deduced from simple physics models.  A given resonance of the instrument can contribute to sound production or sound suppression depending on the degree of its loss to friction.  It is possible to study frictional dissipation  empirically.  By and large, higher frequency motions lose a larger fraction of their energy to heat than lower ones.  Furthermore, the sound radiation efficiency of soundboard vibrations grows with increasing frequency.  Together, these effects make the higher frequency sounds weak and short-lived.  The total sound is dominated by the response of a handful of  low frequency resonances of the body.  And their properties are sensitive to the detailed design of the instrument.

The vibrations of a wood soundboard depend on its internal properties and the applied forces but not particularly on the surrounding air.  In contrast, drums have long been of intellectual interest in physics precisely because the surrounding air and produced sound have a strong impact on the actual head motion.

In ref.~\cite{acoustics}, a variety of lines of inquiry show that heat loss is relatively small in any part of banjo sound production.  This means that dissipation-free, linear equations govern most of what's going on.  The equations imply that  the drum head has increasingly more resonances than a wooden soundboard as frequency increases and that the radiation efficiency decreases with increasing frequency.  Compared to guitar, together these give a sound spectrum that is much richer at high frequencies, and those high frequency components have sustain that is comparable to the low frequency ones.
 
Several observations come together to support this perspective on dissipation in the banjo and its contrast to the guitar.  The simplest evidence presented in ref.~\cite{acoustics} is the success of the analytic and numerical modeling in reproducing measured behavior with little or no energy loss besides energy carried away by sound.  Somewhat more subtle are analyses of pluck sounds into mode frequencies and lifetimes.  On guitar, recognizable string modes have much longer lifetimes than body modes.  This gives strings the time to dissipate energy themselves without most of it proceeding to sound production.  On banjo, the string resonant frequencies die off on times comparable to body modes, suggesting that's how strings lose most of their energy, i.e., by transferring it to the body.

To go further, we turn to a particular measure of how plucked strings get the soundboard moving.

\section{bridge admittance}

Imagine plucks on a banjo and a comparable size guitar, strung and tuned identically.  Following a long and successful tradition in the study of guitars and violins, we focus on precisely what happens when the vibrating string applies a force to the bridge.

The characterization of bridge motion in response to string force is {\it admittance.}  Admittance is the velocity at a point divided by the applied force that produces it.  For vibrations in the context of linear physics, we can consider forces,  motions, and admittance as functions of frequency.  In general, admittance is a complex quantity, with its phase reflecting the phase difference between the driving force and the produced velocity.  
Viewed as a function of frequency, the admittance at the bridge involves all of the subsequent physics --- not just the bridge itself but also the whole head and the surrounding air.  (Admittance is precisely the inverse of impedance.)

Energy considerations offer another perspective on the interpretation and importance of admittance.  The up-and-down motions of strings apply oscillating forces to the bridge.  However, energy is transferred from a string to the bridge only to the extent that the contact point moves.  Otherwise, there is no energy transfer.  So admittance is a direct measure of how much of the pluck energy gets to the soundboard.

\begin{figure}[h!]
\includegraphics[width=4.5in]{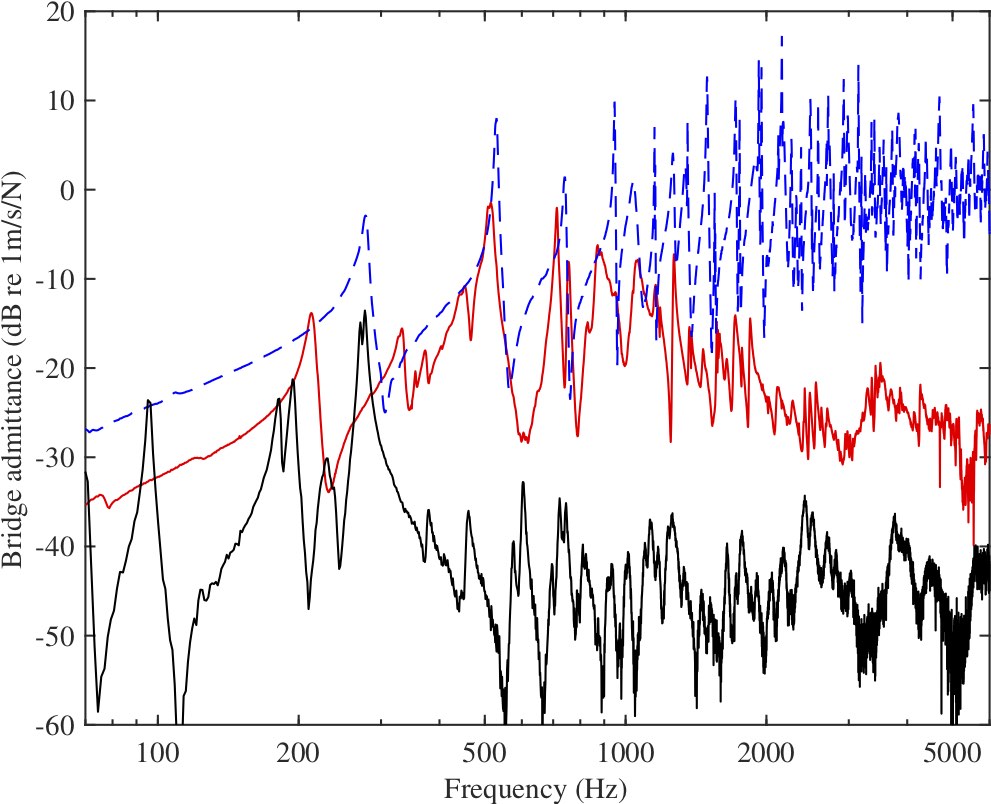}
\caption{Measured bridge admittance at the location of the 1st string slot: blue is the head without bridge \& strings, red is with bridge and damped strings, and black is guitar with damped strings.}
\end{figure}

Fig.~1 shows measured admittance of a Deering Eagle II banjo and a Martin Woodhouse steel string acoustic guitar.  The guitar is in black.  The blue curve is for the banjo { \it without} bridge and strings,  forced on the bare head at the location where the bridge ordinarily goes.  The banjo with bridge and damped strings is in red.  The vertical applied force and measured velocity are at the location of the first string on the bridge.

The contrast of guitar in black with the bare drum head in blue illustrates what's been said so far.  The drumhead soundboard is louder overall --- because it is much lighter.  And it is increasingly more responsive at higher frequencies because there are more and more resonant modes.  Specifically, aside from their lowest several modes, wood soundboards have resonances that, on average, are equally spaced in frequency.  On a drum head, the density of resonances increases proportional to the frequency.

Add strings and bridge (and tailpiece) to the drumhead, and the behavior changes radically.

The up-down-up-down-up-down aspect of the curves initially obscures some underlying trends.  Imagine a running average, smoothing out the curves.  The guitar admittance is essentially flat.  The bare drum head curve  rises.  Adding strings, bridge, and tailpiece lowers the drum head admittance for low frequencies and lowers it again at high frequencies.   This produces a broad intermediate region of strong response, which, in acoustics, is called a ``formant."

Ref.~\cite{acoustics} reports an explicit calculation for a rectangular banjo with an ideal membrane for a head, damped strings, bridge, and break angle to tailpiece.  (Rectangular geometry has long been a favorite of physicists because analytic and computer calculations are far simpler than the corresponding ones for a circular head; they do still sound like banjos.\cite{squared-eel})  The well-isolated lowest modes are sensitive to the geometry, but the smoothed trends are just like in Fig.~1.  

A very general perspective, pioneered by Eugen Skudrzyk\cite{skudrzyk}, predicts the smooth curves and identifies the crucial elements.  The smooth curves reflect the following observations.  The individual peaks of the actual admittance occur at frequencies for which the waves reflected off the soundboard boundary return to the point of admittance in phase with the initial disturbance and essentially double it.  The intervening troughs occur at frequencies for which the reflections return $180^{\text{o}}$ out of phase and cancel it.  The average is the same as if there were no reflected wave coming back at all.  All of the head motion is outgoing waves initiated by the bridge motion.  That is what would happen if the soundboard were infinite in extent.  So, the smooth curve is that of an infinite head --- or, equivalently, any shaped finite head with a non-reflective edge.  Therefore, the smoothed curve for any size and shape head with the same tension and density would be the same.  (Calculations for an infinite plane are even easier than for rectangles.)  

The solid red line in Fig.~2 is derived using the membrane wave equation.  That equation is simply $F=ma$ for every tiny patch of the head.  The mass of each patch is proportional to its area.  The force on each patch is the vertical (i.e., perpendicular to the head) component of the tension of the adjoining patches.  It is the same differential equation for every patch.  How a patch moves depends on what its neighbors are doing.  The admittance is the motion of the particular point where an external force is applied.

Adding the bridge and (damped) strings changes the head equation of motion {\it only at the location of the bridge}, i.e., where it contacts the head.  The simplest approximation to a bridge is to model it as a point mass at the location where the admittance is evaluated.  So there is extra inertia at that point and extra forces --- in addition to the applied force and the uniform membrane wave forces.

Even though banjo bridges are light, e.g., 2.2 gm, their mass is significant compared to the nearby head, which might be 11 gm total for the whole $11''$ D circle.  An additional mass at the point of admittance suppresses the high frequencies.  Larger mass gives greater suppression.  One can say that bridge mass produces a low-pass filter, with the transition frequency  decreasing with increasing mass.

The origin of the extra forces at the admittance point due to strings and head is a bit more complicated.  At equilibrium, those forces sum to zero.  However, when the bridge moves away from its equilibrium position, there is a net return force due to strings and head that is {\it in addition} to what is supplied by the string and head {\it waves.}  For small motions, it is a Hooke's Law force, i.e., linear in the displacement.  It has four sources, and each is under the builder's and player's control --- at least to some extent.   The string break angle is non-zero to keep the strings in contact with the bridge as they vibrate.  In practice, it is typically between $5^{\text{o}}$ and $15^{\text{o}}$ at equilibrium.  1) The bridge motion changes the  break angle.  That alters the vertical component of the string tension on the bridge.   And 2) there is also a change in the string tension --- a higher tension when the bridge displacement is positive --- because the string must stretch when the bridge moves up.  The increased-tension contribution to the Hooke's Law return force is proportional to the square of the sine of the break angle and the stretchiness of the strings (the longitudinal consequence of the string's Young's modulus).   Bridge motion also produces 3) an analogous change in the break angle of the head at the feet of the bridge (albeit less apparent to casual observation) and 4) requires some stretching of the head, which increases its tension.

These four effects add a springlike return force right at the admittance point, which is {\it in addition} to the return force provided everywhere on the head by its tension.  In ref.~\cite{acoustics}, this is called the ``stiffness."  Stiffness is simple to implement in the Skudrzyk calculation.  It produces a suppression of admittance for low frequencies, which is increasingly effective as the break angle increases.  Equivalently, this is a high-pass filter.

The effects of the extra mass and\ return force at the bridge combine to yield a central frequency region of largest admittance --- a formant ---  and the center of that region reflects the  balance of the two --- as illustrated in Fig.~2 by the dashed blue curve.   Using realistic parameter values, this compares well to the corresponding measurement in Fig.~1, shown in red. 
The significant discrepancies above 3 kHz are due to the additional formants discussed
in the next section, {\small \bf \S IV}.
\begin{figure}[h!]
\includegraphics[width=4.5in]{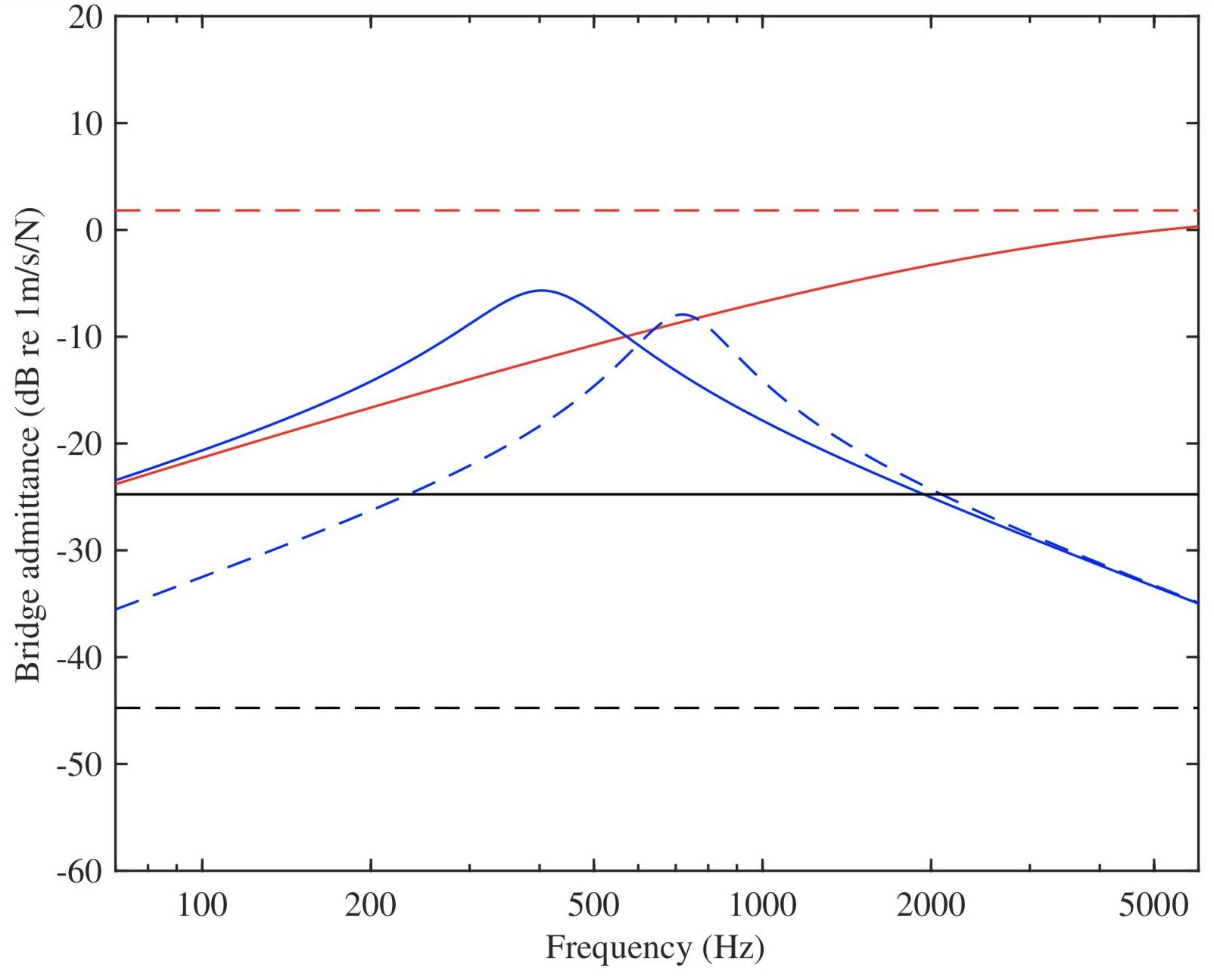}
\caption{\footnotesize Skrudrzyk admittance calculations --- 
red solid: head alone; red dashed: high frequency limit of head alone; blue solid: head with a 1.5 g bridge mass; blue dashed: head with a 1.5 gm bridge, realistic stiffness, and a small amount of damping; black solid: wood soundboard with mass per unit area matching the head; black dashed: wood soundboard with mass increased by a factor 10 (realistic for a thin piece of wood).}
\end{figure}

The same equations can be described with different words.  We can consider the bridge mass and string/head stiffness as defining a single oscillator.  Its motion is very strongly damped by the energy loss to outgoing waves in the head.  Very strong damping implies a very short decay time and a very broad width in frequency.   If the damping were independent of frequency, it would reduce the peak frequency in the admittance relative to the undamped oscillator value. However, the rising trend of the head admittance pushes in the other direction, leaving the total, net peak frequency shift subject to the detailed numerical values.

This calculation identifies the physics behind simple modifications that are commonly used to alter the sound of a particular banjo.  For example, consider more mellow.  Here, ``more mellow" means lowering the central frequency of this main formant.  You can increase the mass of the bridge.  And there are several ways to decrease the stiffness.  Bridge height and tailpiece design have obvious impacts on break angle; smaller angle is more mellow.  Lower string and head tension and easier stretchiness each promote mellow.

The bridge is held in place by ``pre-tensioned" strings and head.  At equilibrium, their forces cancel.  But it is sometimes said that the pre-tension enhances the communication of string motion to the head.  Indeed, the discussion above shows that higher pre-tension raises the main formant frequency.  However, this actually happens by suppressing low frequency motion rather than enhancing the highs, as illustrated in Fig.~2. 

\section{formants from bridge flexing}

Careful observation of the Deering Eagle II revealed two additional formants at higher frequencies, i.e., centered around 3.5 kHz and 5 kHz.  The basic admittance measurements reported in ref.~\cite{acoustics} were performed by tapping downward at the $1^{\text{st}}$ string bridge slot.  Two further strategies revealed more about the higher formants' origins.  Tapping downward at the $3^{\text{rd}}$ string slot enhanced the 5 kHz formant, and tapping sideways on the edge of the bridge enhanced the 3.5 kHz.  In those frequency regions,  finite element numerical simulations showed dramatic bridge flexing.  Note that these are not resonant motions of the bridge on its own.  Rather, the resonant motion requires a bridge flexing that fits with head motion at the same frequency.   The head motion is particularly large at the bridge feet while they maintain contact.  Fig.s~3 and 4 show the result of the numerical calculation of  resonances near 3.5 kHz and 5 kHz.  (FE/BE refers to a discretization of the solid objects, i.e., ``finite element," and a continuous fluid calculation of the surrounding air whose specific behavior in any situation is required to match the motion of its solid boundaries, i.e., ``boundary elements.")
\begin{figure}[h!]
\includegraphics[width=4.5in]{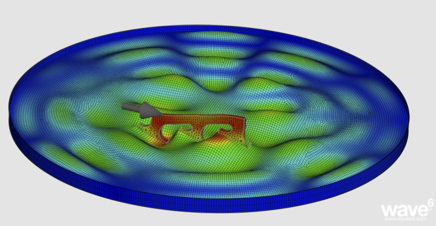}
\caption{FE/BE calculation of resonant bridge/head motion near 3500 Hz.}
\end{figure}
\begin{figure}[h!]
\includegraphics[width=4.5in]{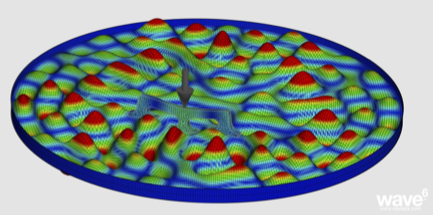}
\caption{FE/BE calculation of resonant bridge/head motion near 5000 Hz.}
\end{figure}

Note that these extreme bridge flexing motions are peculiar to the bridge used in the measurements and numerical calculations.  The bridge in question is three-foot ebony-topped maple, standard equipment on the Deering Eagle II.  Analogous effects occur at different frequencies with different bridge designs, the relevant factors being flexibility and footprint.\cite{bridge-hills}

The speed of sound in hardwood is 10 to 12 times greater than sound in air.  So the wavelengths of longitudinal waves of compression are 10 to 12 times longer.  For relevant frequencies, these are at least 20 times longer than the vertical dimension of a typical bridge.   That means, in terms of sound traveling as compression waves {\it through} the wood, the bridge moves as one, i.e., as a single, rigid piece.  There is no real meaning to it taking a particular path through the bridge.  On the other hand, bridges can flex on length scales comparable to their own geometry, and those frequencies can be much lower than those of bridge-scale compression waves.

\section{some important physics}

\subsection{wave speeds}

Discussing sound production by a soundboard requires establishing a bit about the physics notions of an ideal drum head and an ideal wooden soundboard.

Just as there is a physics model and equation that gives a reasonable, approximate account of thin, taut musical strings, there are well-established physics models for the ideal drum head, known as a ``membrane," and the typical soundboard of a wooden string instrument, known as a ``thin plate."   For the present purposes, members of the international banjo family are instruments whose heads behave, to a first approximation, like ideal membranes.
 
The ideal string is a physics abstraction that does a reasonable job accounting for the behavior of the thin strings of real musical instruments.  The ideal string has a uniform density and a uniform tension that provides the only force that pushes the string back toward its equilibrium position when it is disturbed.  Simple observations confirm that this idealization is a very good approximation for small amplitude vibrations.  One need only note how well it works.  Obvious forces that are neglected are bending stiffness and internal friction.  The long lifetime (in terms of number of cycles) of a plucked string implies that internal friction is not a primary issue.  Bending stiffness makes successive string resonances sharper and sharper relative to the ideal.  This can be observed as a small effect on solid, thick strings.  And that's why they're usually wound  -- to add mass without significant extra stiffness.

The membrane is an ideal 2-dimensional abstraction.  It has a uniform, isotropic tension that provides the only force that pushes the membrane back toward its equilibrium position when it is disturbed.  This specification of mass and force gives rise to wave behavior that is well-understood --- although not previously explored carefully in the context of banjos.  For the present purposes, a banjo is any string instrument whose soundboard can be approximated as a membrane.  In practice, that means that the tension is high enough, head thin enough, and amplitudes small enough that tension dominates over other solid material forces in determining transverse motion.   Inherent stiffness or rigidity are ignored.  The closest approximation to an ideal membrane in the real world is a soap bubble film.    Indeed, an ideal membrane is actually something of a two-dimensional liquid. 

The thin plate is an idealization that is known to work well for wooden soundboards.  The restoring force is a stiffness arising from the stretch and compression inherent in bending a solid object that is extended in two dimensions and thin in the third dimension.  The resulting wave dynamics is quite different from the membrane.  Beating the general description of deformation of an elastic solid into a convenient form for thin plates is a more complicated story than needed for the ideal membrane.  But a very useful result emerges, and it is a standard topic in mechanical engineering education.  In physics education, it typically only appears as an advanced example, applied only to the static equilibrium deformation of a beam under load. 

The resulting wave equations imply that on a membrane, transverse plane waves all travel at the same speed. On a plate, the speed of waves $v$ of definite frequency $f$ is higher for higher frequencies, i.e., $v \propto \sqrt{f}$.

\subsection{static deformation \& low frequency limit}

A constant small force applied at a point along a string produces a small displacement.  That is also true for a plate.  However, for a membrane, the displacement is infinite for any non-zero force applied at a point.  The reason is simple.  Tension $T$  on a 2D surface is a force per unit length.  If the applied force $F$ is spread uniformly over a disk of circumference $2\pi r$, the resulting static deformation requires $F = 2 \pi r T \times$ (the slope of the membrane at $r$).  Hence the slope at the edge of the radius $r$ disk is {\large ${F \over {2 \pi r T}}$}.  That implies that the deformation goes like log\thinspace$r$ as $r\to 0$.  Of course, bridges have finite sized feet.  But one lesson is that small bridge feet can enhance the relevance of forces that were otherwise negligible, e.g., bending stiffness near the bridge.  Measurements as a function of $r$ and calculations for a rectangular membrane and square footprint are shown in Fig.~5.

\begin{figure}[h!]
\includegraphics[width=4.5in]{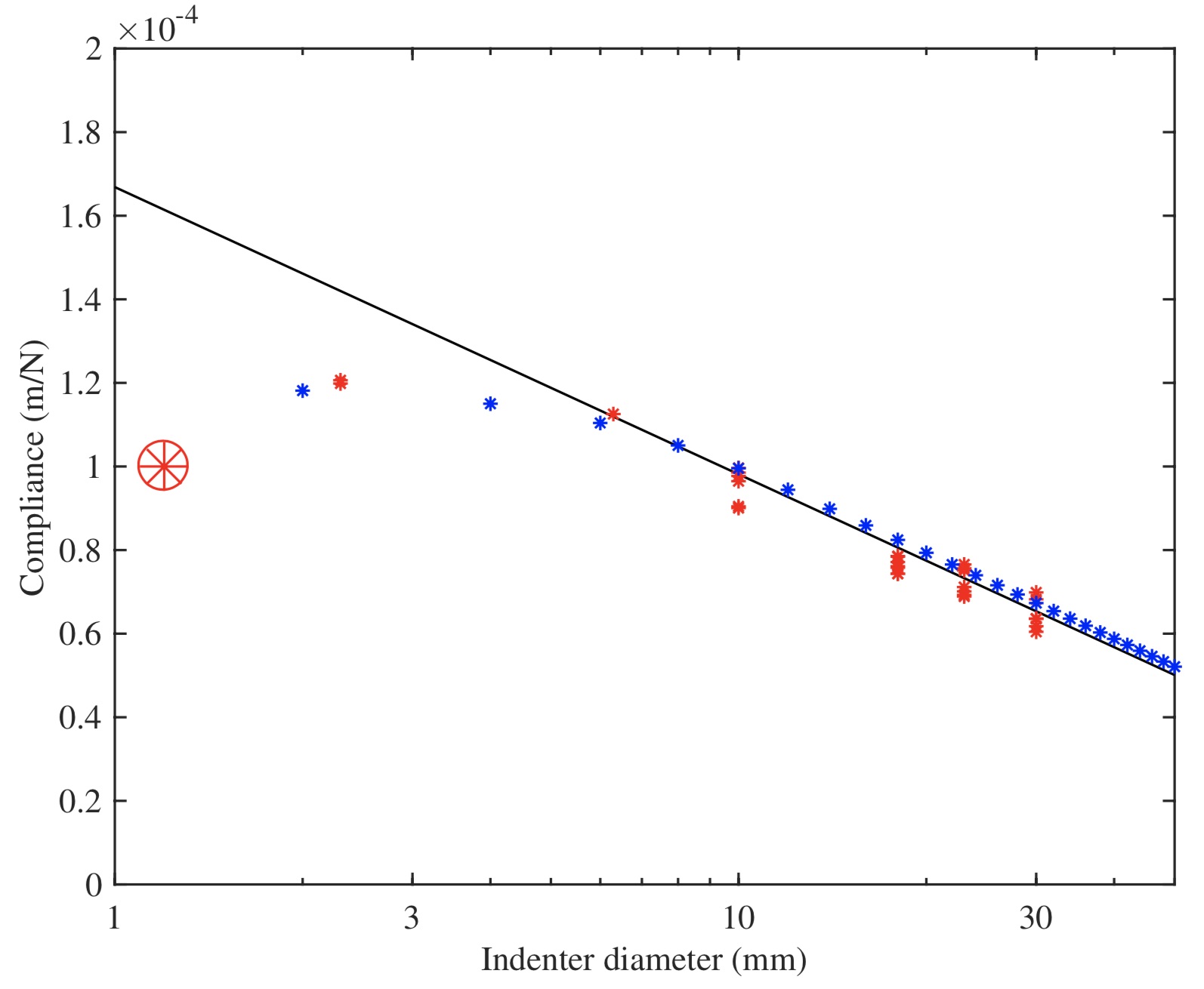}
\caption{Compliance (static displacement divided by total applied force) {\it versus} diameter of the forcing disk. Red stars: measured static compliance at the centre of the head of the Deering banjo, for different diameters of rigid indenter. (See photo on page 1.)  Solid line is the prediction from ideal membrane theory.  Blue stars: values deduced for a square loading region at the centre of a square membrane, with a cutoff frequency 10 kHz.  Large red symbol: DC compliance deduced from the $\omega \to 0$ measured admittance using a hammer with a very small area of contact.}
\end{figure}

The relatively extreme bending of the head at the feet of the bridge has a dynamical effect, beyond the static displacement.  Very high frequency drum-head modes can impact the low frequency motion of the bridge, even in an analysis restricted to modeling linear forces.  Here is an outline of how that comes about:

Each membrane resonant mode behaves like a forced harmonic oscillator.  Each has a resonance peak.  Below its peak, the amplitude of each one approaches a finite value as the forcing frequency goes to zero.  That value is its static equilibrium displacement.  The full membrane is just the sum of its modes.  The dynamical admittance measurement, evaluated well below the frequency of the lowest mode is essentially the static compliance.  The large red wagon wheel in Fig.~5 is an estimate of that value from admittance measurements of a real head excited by a tiny hammer.  So the static compliance at the bridge is made finite by some combination of effects that are not part of the simple physics that otherwise give a first approximation to everything that goes into banjo sound production.  Contributors certainly include other forces.  In addition, some connection to the infinity of modes as $f \to \infty$ may be relevant because it is the sum over all those modes that gives the naively infinite point compliance.  And all these effects have significant impact on the structure of the lowest frequency modes of the finite banjo head. It is unusual in acoustics and much of physics that the lowest frequency, longest wavelength phenomena are sensitive to very short distance physics.  In this case, stiffness at short distances enters DC compliance because of the sharp bending localized right at the feet of the bridge.  That sharp bending also means that short wavelength drum modes can have a significant impact on low frequency motions --- and it happens because of the underlying singular behavior of the point compliance.

These various effects produce an extreme sensitivity of frequency and lifetime of low frequency modes to small variations of parameters, as illustrated in $\S3.4$ of the second paper of ref.~\cite{acoustics}.  There, the ``toy" calculation studies the lowest two modes, modeled as damped harmonic oscillators with their own mass, spring constant, and damping, as they are impacted by the contribution to the static compliance due to all other, higher frequency modes.  The resulting frequencies and lifetimes are shown to be very sensitive functions of that compliance.  The take-away is that the details of the lowest modes are {\it very} sensitive to bridge physics and the static compliance.

\subsection{sound radiation efficiency}

When a surface vibrates at a given frequency, a relevant issue is the wavelength on the surface {\it versus} the wavelength in air.  If the surface wavelength is smaller than that in air, neighboring pieces on the surface tend to cancel each other's contribution to the sound.  For an infinite surface, the cancellation is complete.  With a plate, there is a critical frequency at which the two wavelengths match (i.e., where the wave velocities are equal).  Above that frequency, the plate waves are longer than air waves, and sound radiation is far more effective than below it.

For any physically reasonable banjo head, the speed common to all waves is less than the speed of sound in air.  Hence, an infinite head would not radiate sound at all.  In more practical terms, the radiation of sound is intimately connected to the finite size of the head.  This has an important, simple consequence.  Radiation efficiency decreases with increasing frequency for a given size of head.  That's because for high frequency (and shorter wave length) the head is effectively closer to infinite as measured in units of the wavelength.  This means that the high harmonics in a single pluck sound do not die off quickly relative to the rest.  In fact, all harmonics have comparable sustain.

\section{body sound in each pluck}

The simplest picture of stringed instruments is that strings drive the soundboard at their 
near-harmonic frequencies.  For various reasons, the string amplitudes decay, and the sound decays as well.  However, a pluck involves a sudden initiation.  That actually excites all of the modes of the soundboard/body, not just the resonant frequencies of the string.  On wood-topped instruments, most of this energy goes into heat, and the soundboard/body modes die away very quickly relative to the string modes.

Fig.~6 shows spectrograms of the  same note plucked on a banjo (above) and a guitar (below).  The tall, thin lines are related directly to the string modes.  The other stuff at the bottom comes from the body modes.
\begin{figure}[h!]
\includegraphics[width=3.8in]{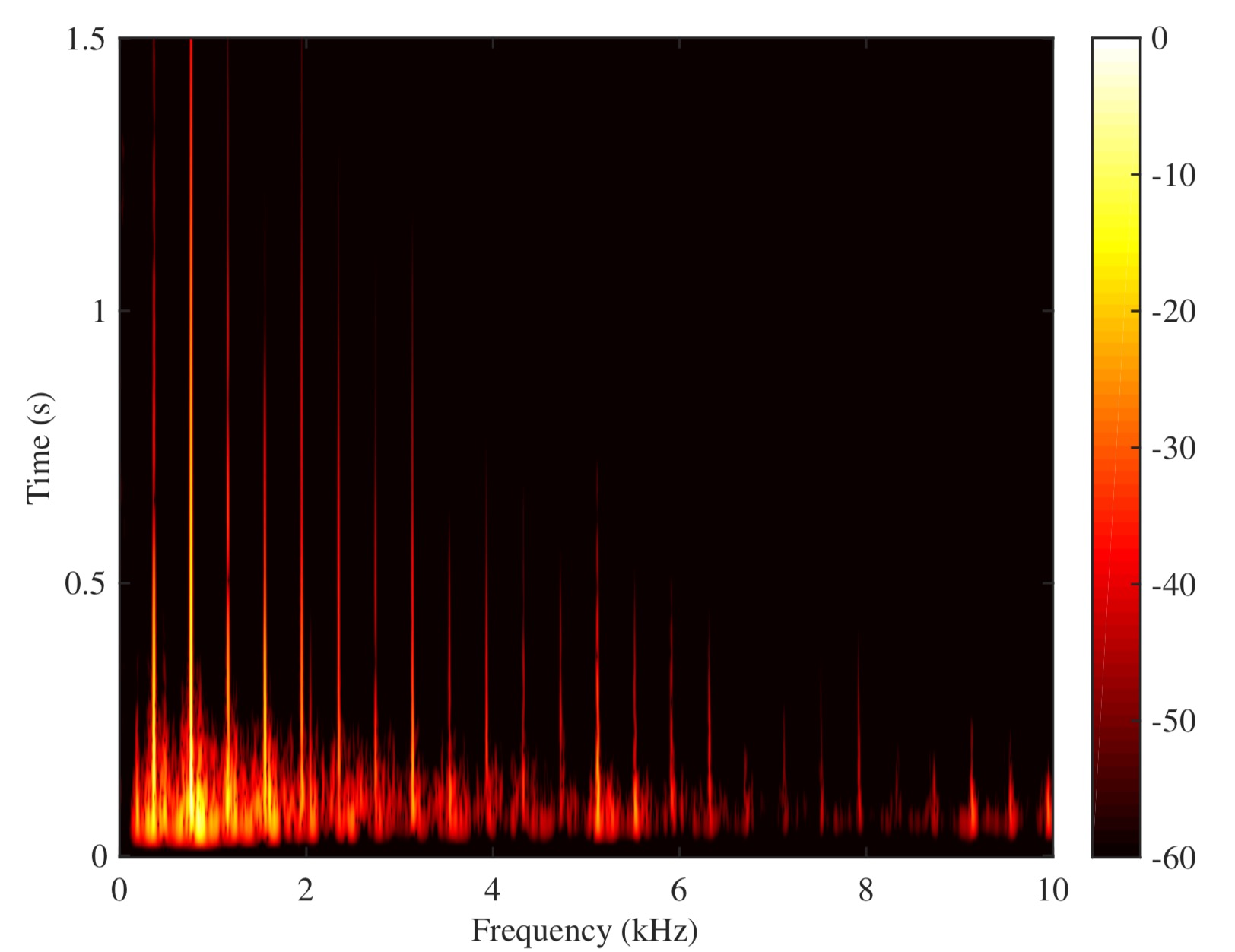}
\includegraphics[width=3.8in]{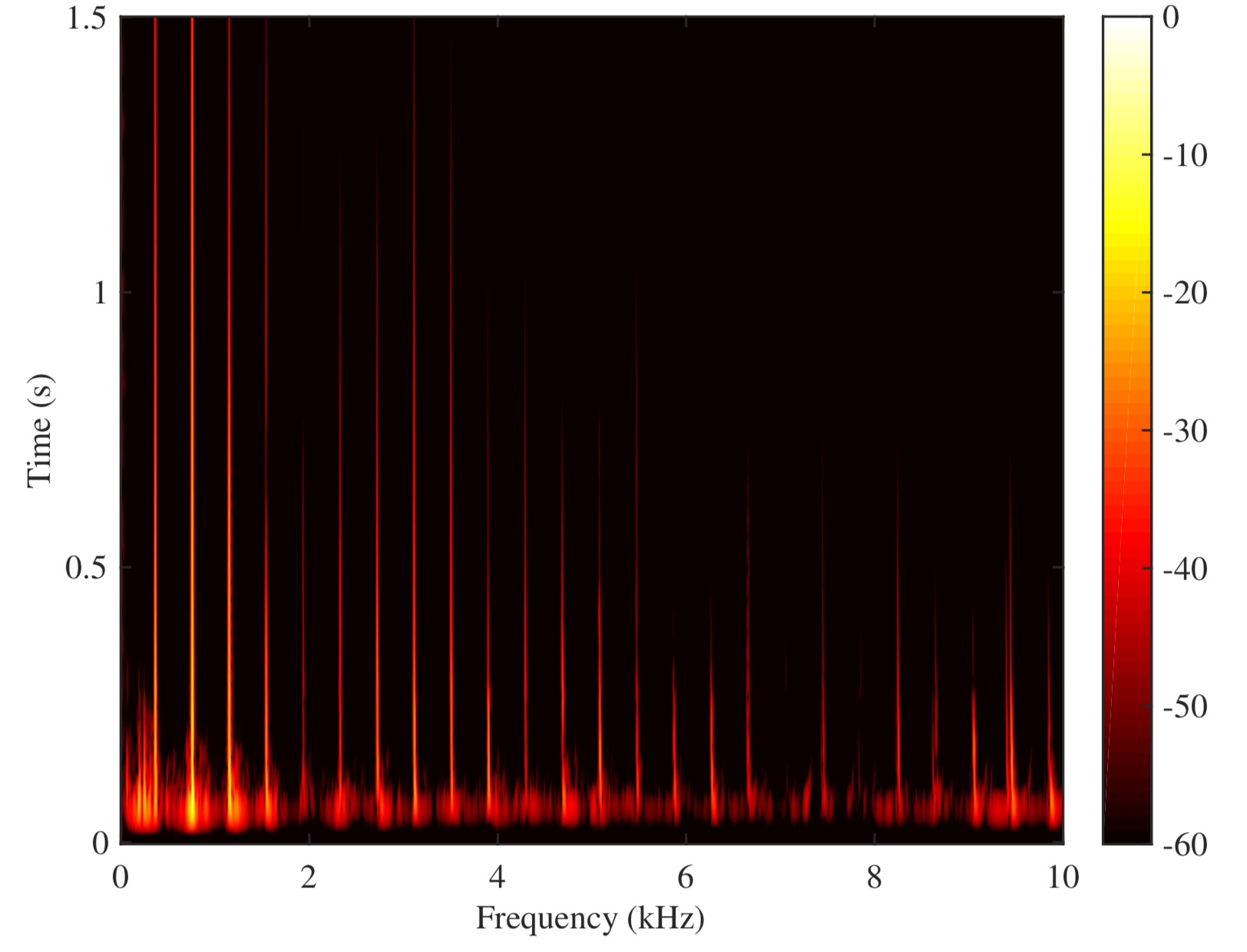}
\caption{Spectrograms of a single pluck on the $1^{\text{st}}$ string (G4 -- 392 Hz) on a banjo (top) and guitar (bottom).}
\end{figure}

Three things are different on the banjo.  1) Almost all of the energy goes into sound.  2) String and head/body mode decay times are not as dissimilar as with guitar, primarily because the strings dump their energy into the head more readily.  And 3) banjo music generally consists of a steady stream of closely spaced notes.  So, the decaying head/body sounds survive for a significant fraction of each note.  Note that four notes per second, i.e., spaced by 0.25 s, is not particularly fast for banjo music.  Hence, head/body modes survive for much of the time between notes.

\section{synthesized sounds from physical modeling}

The sound files of ref.~\cite{euphonics}, \S5.5, offer a way to listen to a generic synthesis generated by the linear, physical modeling.  You can explore how that sound changes when the various parameter values are changed one at a time.

One important caveat is that all generated sound files give audio realizations of the motion of the bridge and not the actual produced sound somewhere in space.  This has several advantages.  It is precise and concise.  Note that the bridge motion in response to some pluck or tap reflects not only the physics of the bridge and soundboard but also that of the strings, the surrounding air, and all their mutual interactions.  Bridge motion is crucial in translating string vibration into produced sound.  And calculating the impact of changes in various model parameters is a feasible task.  In contrast, the sound at some point in space depends on the relation of that point to the instrument and also to the space they are in, i.e., the room and everything else in it.  Furthermore, that is a calculationally intensive task that would be a practical obstacle to comparing the results for ranges of the various parameters.  

To get the bridge motion ``sound" to be a bit closer to real instruments' sound, all the synthesized files include a smooth, slowly varying, monotonic filter that boosts the highs slightly: $\times f^{0.4}$ for frequency $f$.

By and large, physically motivated parameter values give a plausibly banjo-like sound.  And sound variations produced by varying the parameters track changes that are commonly experienced with real banjos.

\section{conclusion}

The essential feature of banjo is the drum head soundboard.  Ref.~\cite{acoustics} presents efforts to understand how that works, i.e., to get a first approximation of the linear behavior of the basic parts.  Membrane physics requires that the bridge and tailpiece (through the break angle) play essential roles and are not merely passive conveyors of string vibration to the membrane, as is the case with the bridge and saddle on a guitar.  This note sketched some of the results and underlying physics.  Many features are in stark contrast to the response of a similarly strung guitar.  The compelling details are in ref.~\cite{acoustics}.  Even a perusal of the figures and their captions and a listen to the sound files would give a more complete picture than was presented here.  For the most part, measurements were made on only one banjo with one bridge.  But its features were successfully modeled in considerable detail.  The principle input parameters were simply the design choices of strings, head, and bridge.  And the rectangular banjo model produced a very similar basic behavior.  We conclude that the physics presented here gives a complete first approximation to banjo acoustics.

Even though ref.~\cite{acoustics} studied only one bridge and one banjo, the comparison of measurements and calculations provides a compelling picture.  That offers many suggestions of how modifications of the instrument will change the sound.  For example, bridge flexing is associated with the production of high frequency formants, i.e., regions of enhanced sounds.  Hence, bridges with different flex characteristics are expected to sound different.\cite{bridge-hills}

\bigskip

\centerline{\bf ADDENDUM: HEAD TAP TUNING}

\bigskip

Ref.~\cite{acoustics} contains many observations and perspectives that did not make it into this summary.  But something of possible practical interest to players concerns tap tuning the head for tension adjustment.  The ultimate tensioning advice is to adjust the tension to what sounds best to you.  But there are many reasons to use a method that involves less trial \& error.  One might want a new head to sound like the previous one.  Or one might aspire to the sound of someone else's banjo.  So a common recommendation is to tap and adjust the tension to a specified pitch.  On an $11''$ mylar head, bluegrass tight is G$^\#$ or A.  The problem is that not everyone can clearly identify the tap pitch.  Some electronic tuners will catch it, and some won't.

A clue was identified by author and banjo novice J.W., when he had to mount a new head to replace the torn original.  He recognized a connection with features of the many graphs of spectra and images of modes that we generated.  (E.g., consider 
ref.~\cite{acoustics}, second paper, Fig.~7.)  }For tap tuning, one is instructed to tap or scratch, with a very small area of contact, near the rim.\cite{huber}  J.W. found that damping the central region of the head, e.g., by damping the bridge, enhanced the desired effect.

Highly localized tapping near the edge along with damping the interior excites a motion characterized by short wavelengths along the rim and extending only one wavelength or so inward in the radial direction.  These are the whispering gallery modes studied by Rayleigh\cite{whisper} to account for the whispering gallery of St.~Paul's Cathedral in London.

Because it's a sharp tap or scratch near the rim, the frequency spectrum consists of many narrow peaks, going up to substantially high frequencies.  However, the perceived tap pitch is not one of them.  The key to understanding what's going on is to look at the {\it differences} between successive frequency peaks.  Those numbers rapidly approach a single value (asymptotic from above).  Although there is no peak whose frequency is that difference value, the combination (``superposition") of the motions represented by those peaks is approximately periodic with a frequency given by the difference.  (For this to make sense, it is essential to remember that the spectrum analysis identifies {\it sinusoidal} Fourier components and not simply all motions that repeat with a fixed period.)

Depending on hearing acuity, brain function, and how close the actual frequency differences are to multiples of a common factor, people tend to identify the pitch of such a complex sound as that associated with the common difference.  In psychoacoustics, this is called the phenomenon of the ``missing fundamental."

In some situations, hearing the missing fundamental is absolutely unambiguous.  For almost any banjo and tuning, a pluck in the lowest half-octave on the $4^{\text{th}}$ string produces essentially no sinusoidal frequency component equal to the pitch.  The note is heard nevertheless.  In fact, the timbre is not noticeably different from the slightly higher notes on that string whose frequency spectra do, indeed, include the fundamental. 

In the whispering gallery mode description, the difference frequency described above is the fundamental frequency of the whisper.

Here is a specific example of a very tight head on a Deering Sierra, tapped at the edge with a pencil eraser, damped at the bridge, and recorded very nearby:
\href{http://www.its.caltech.edu/~politzer/pickers-guide/whisper-1-just-one.mp3}{click here} or go to \href{http://www.its.caltech.edu/~politzer/pickers-guide/whisper-1-just-one.mp3}{http://www.its.caltech.edu/\url{~}politzer/pickers-guide/whisper-1-just-one.mp3} .

Fig.~7 is a spectrogram of a series of such taps.  The nearly equal frequency spacing (i.e., in the vertical direction) of intensity peaks is obvious.

\begin{figure}[h!]
\includegraphics[width=6.5in]{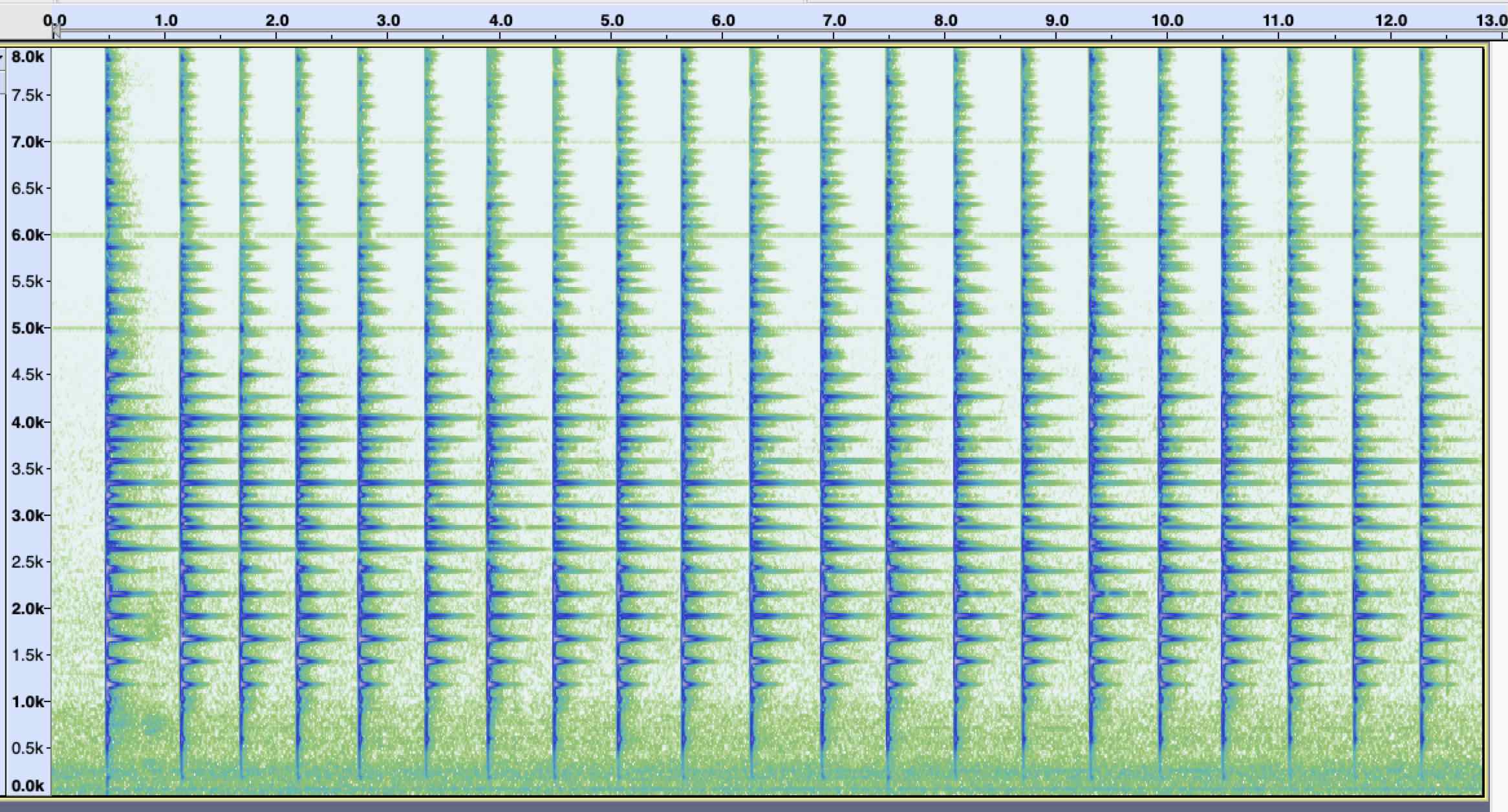}
\caption{spectrogram of a sequence of 21 taps}
\end{figure}

Using the spectrogram, the common spacing works out to about 237 Hz, which is an A\#.  Hearing-challenged author D.P. had a hard time at first to associate the tap sound with a string pluck pitch.  However, in retrospect, it's a pretty good match.  As usual, the actual octave is ambiguous to some listeners, e.g., it could be 474 Hz.

Fig.~8 gives the computed spectrum of the twenty one taps in Fig.~7.  Indeed, 237 Hz or 474 Hz are nowhere to be seen.

\begin{figure}[h!]
\includegraphics[width=6.5in]{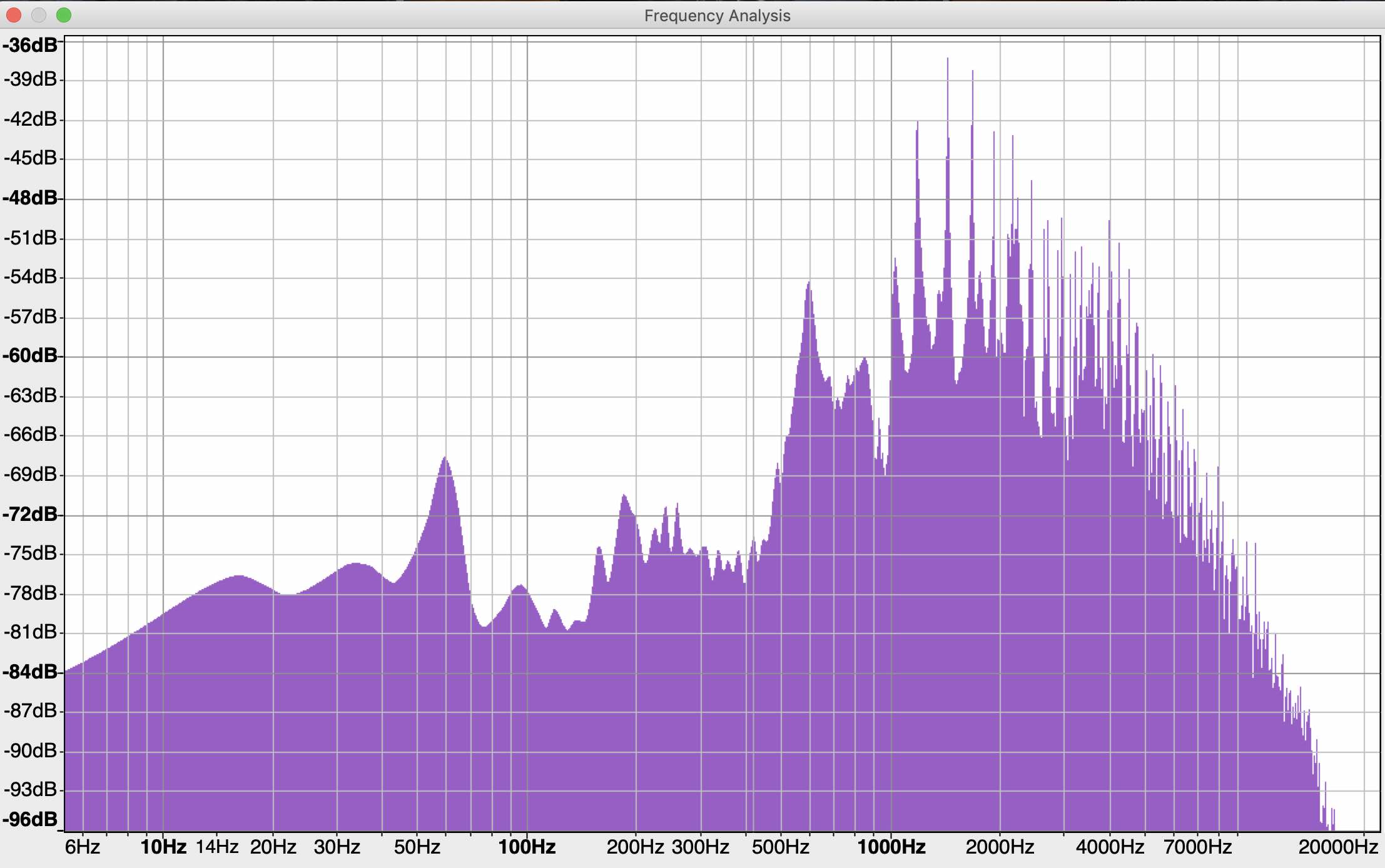}
\caption{spectrum of the sound of the 21 tap sequence of Fig.~7}
\end{figure}

\begin{figure}[h!]
\includegraphics[width=6.5in]{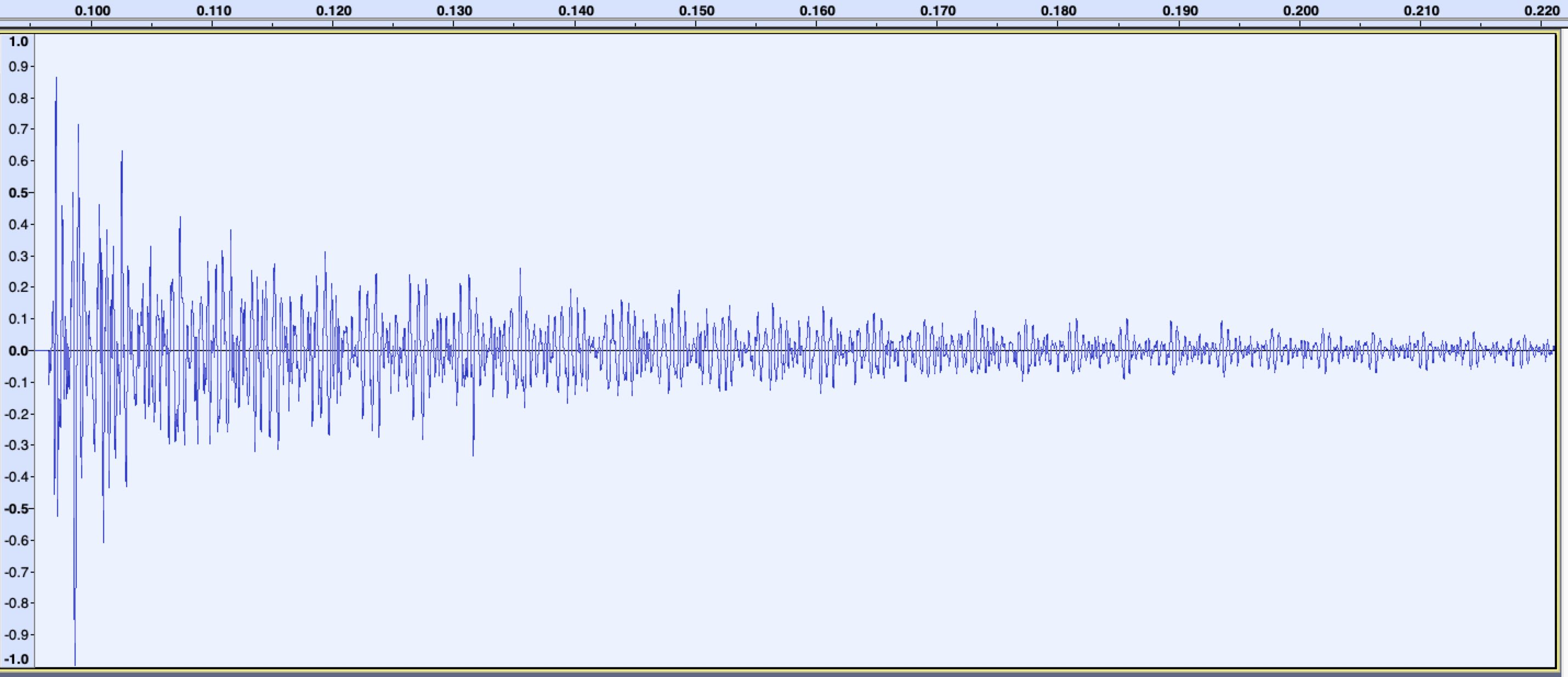}
\caption{waveform of a single tap}
\end{figure}

\begin{figure}[h!]
\includegraphics[width=6.5in]{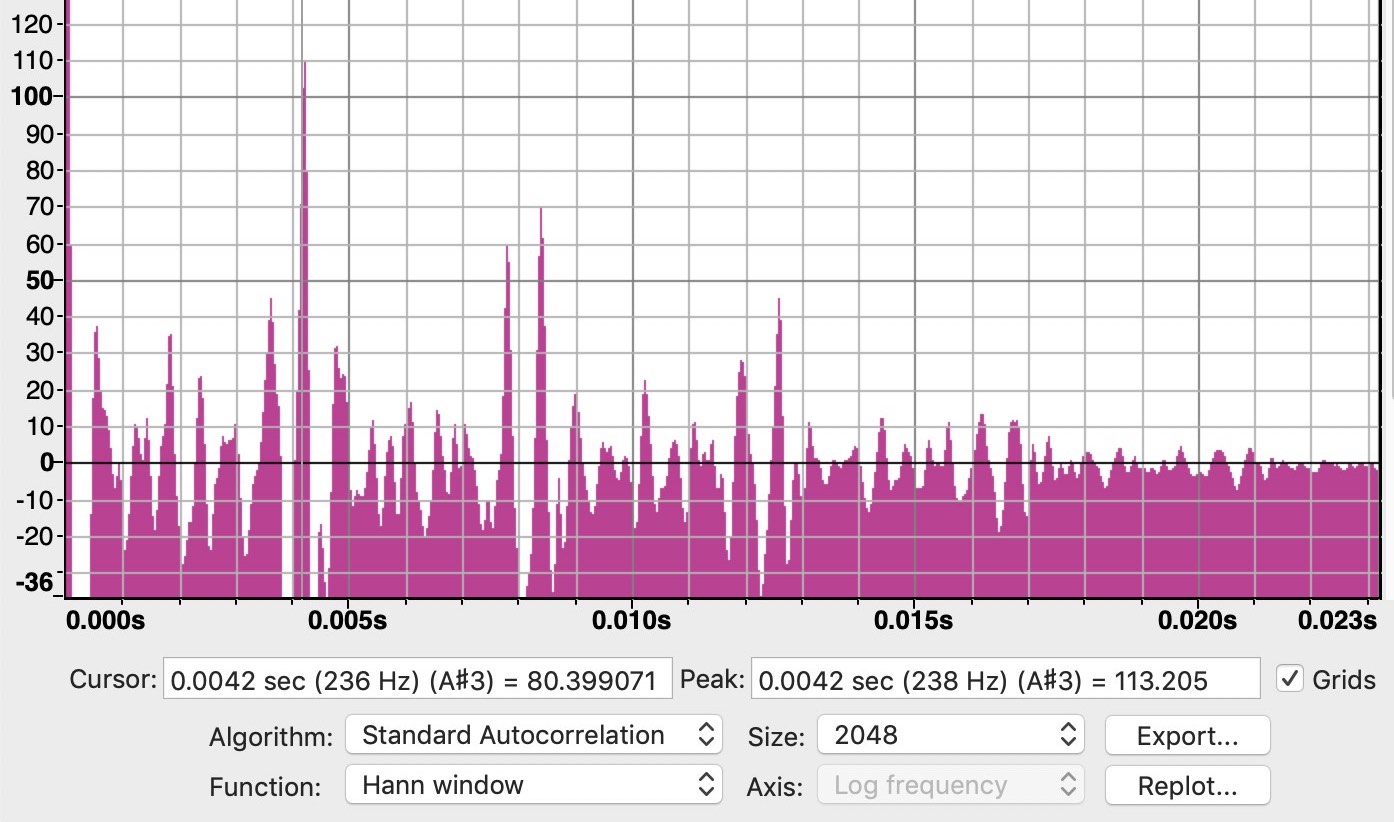}
\caption{autocorrelation of a single tap}
\end{figure}

The waveform of a single tap is shown in Fig.~9.  A rough periodicity can be estimated from from the graph by counting 29 cycles in 0.121 s.  That corresponds to 240 Hz.  Autocorrelation is a well-defined mathematical calculation that can be performed on a signal.  In cases of complicated sources, the location of a single high peak of the autocorrelation generally agrees with the brain's attempt to identify a single pitch.  Audacity's calculation of the autocorrelation of the tap from Fig.~9 is shown in Fig.~10.  In the Audacity Frequency
Analysis Autocorrelation window, the entry called ``{\tt Peak:}" locates the highest peak at 

\noindent 0.0042 s, corresponding to 238 Hz (an A\#3).



\begin{thebibliography}{99}
\bibitem{coupled}For example, subtle but discernible linear physics effects include string polarization\cite{euphonics} and sympathetic coupling of harmonics between different strings, which is considerably stronger on banjo than guitar.\cite{sympathetic}
\end{thebibliography}
\end{document}